\documentclass[
 reprint, pra, aps,
superscriptaddress,
 amsmath,amssymb,
 floatfix,
 doublecolumn,
 showkeys,
 linenumbers
]{revtex4-2}

\usepackage{graphicx}
\usepackage{bm}

\begin{document}
\nolinenumbers
\title{Decision-making in light-trapped slime molds involves active mechanical processes}

\author{Lisa Schick}
\affiliation{%
 Technical University of Munich, TUM School of Natural Sciences, Department of Bioscience, Center for Protein Assemblies (CPA), Germany
}
\author{Emily Eichenlaub}
\affiliation{%
 Technical University of Munich, TUM School of Natural Sciences, Department of Bioscience, Center for Protein Assemblies (CPA), Germany
}
\author{Fabian Drexel}
\affiliation{%
    Technical University of Munich, TUM School of Natural Sciences, Department of Bioscience, Center for Protein Assemblies (CPA), Germany
}
\author{Alexander Mayer}
\affiliation{%
    Department of Mathematics, UCLA, Los Angeles, California, USA
}
\author{Siyu Chen}
\affiliation{%
    Max Planck Institute for Dynamics and Self-Organization, Göttingen, Germany
}
\author{Marcus Roper}
\affiliation{%
    Department of Mathematics, UCLA, Los Angeles, California, USA
}
\author{Karen Alim}%
 \email{k.alim@tum.de}
\affiliation{%
 Technical University of Munich,  TUM School of Natural Sciences, Department of Bioscience, Center for Protein Assemblies (CPA), Germany
}%

\date{\today}
\begin{abstract}
\noindent Decision-making is the process of selecting an action among alternatives, allowing biological and artificial systems to navigate complex environments and optimize behavior. While neural systems rely on neuron-based sensory processing and evaluation, decision-making also occurs in organisms without a centralized organizing unit, such as the unicellular slime mold \textit{Physarum polycephalum}. Unlike neural systems, \textit{P.~polycephalum} relies on rhythmic peristaltic contractions to drive internal flows and redistribute mass, allowing it to adapt to its environment. However, while previous studies have focused on the outcomes of these decisions, the underlying mechanical principles that govern this mass relocation remain unknown.
Here, we investigate the exploration process of \textit{P.~polycephalum} confined by blue light into polygonal shapes up to its escape. While the escape occurs along the longest axis of the polygones, independent of confinement shape, the exploration process prior to escape extends protrusions almost everywhere around a shape boundary. We find protrusions to align with the direction of peristaltic contraction waves driving mass relocation. Mapping out contraction modes during exploration in detail we observe an ongoing switching between different dominant principle contraction modes. Only over the course of time does the organism ultimately settle on the contraction mode most efficient for transport, which coincides with the escape. Thus, we find that only harsh environmental confinement triggers optimal behaviour which is reached by long time re-organization of the flow patterns. Our findings provide insights into the mechanics of decision-making in non-neuronal organisms, shedding light on how decentralized systems process environmental constraints to drive adaptive behavior.
\end {abstract}


\maketitle
\thispagestyle{empty}

\section*{Introduction}
\noindent Biological systems constantly face decisions, such as where and how to allocate resources, when to act, or which direction to move~\cite{huang_Model_2021}. Each of these choices arises from a decision-making process - the selection of one option from a set of alternatives based on internal states and external cues~\cite{huang_Model_2021}. This process is a fundamental feature of both biological and artificial systems, enabling adaptive behavior and flexible responses to changing conditions. Although traditionally associated with neural processing in complex organisms~\citep{samaha_Spontaneous_2020, hayden_Neuronal_2011, briggman_Optical_2005, groschner_Dendritic_2018}, decision-making is increasingly recognized as a general principle that also applies to ``simple'' non-neuronal organisms~\citep{huang_Model_2021, severino_Plants_2021, fukasawa_Ecological_2020}.\\

\noindent Among non-neural organisms, plasmodial networks of the unicellular but multinucleated slime mold \textit{Physarum polycephalum} stand out for their ability to solve complex tasks through decentralized decision-making. It consistently selects behaviorally optimal solutions, such as finding the shortest path through a maze~\cite{nakagaki_Smart_2001, tero_Physarum_2006}, optimizing its foraging strategy and diet~\cite{reid_Decisionmaking_2016, latty_Food_2009, dussutour_Amoeboid_2010, latty_Speed_2011} or avoiding harmful light~\cite{hato_Phototaxis_1976, nakagaki_Action_1996, bauerle_Living_2020, chen_Network_2023, schreckenbach_Bluelight_1981} while still exploring food options~\cite{nakagaki_MinimumRisk_2007,  latty_Food_2010}. These behaviors suggest that \textit{P.~polycephalum} integrates environmental cues to optimize its responses.
While most studies focus on the outcomes of these decisions like growth direction or mass redistribution, the question persists of how the decision of where to relocate mass is initially formed.\\

\noindent Once a decision is made, mass is relocated by forming protrusions in the chosen direction through local softening of the cytoskeleton~\cite{oettmeier_Form_2018}. 
In \textit{P.~polycephalum}, migration is driven by intracellular flows generated by rhythmic contractions of the actomyosin cortex~\cite{matsumoto_Locomotive_2008, alim_Random_2013}. These contractions create a peristaltic wave aligned with the longest network body axis that drives shuttle-streaming of the viscous endoplasm inside the tubular structure with a period of around 120~s~\cite{kamiya_Physical_1981, wohlfarth-bottermann_Oscillatory_1979}, with net flow biased toward the front by asymmetries in amplitude and timing~\cite{oettmeier_Lumped_2019, matsumoto_Locomotive_2008}. Pressure gradients arise from coordinated contractions and local differences in cortical stiffness, modulating the flow and shaping protrusions~\cite{ueda_Mathematical_2011, oettmeier_Form_2018, matsumoto_Locomotive_2008}. As the organism advances, its protrusions reorganize into a network of tubes, with veins reinforcing when aligned with the flow and pruning otherwise~\cite{rodiek_Patterns_2015, baumgarten_Dynamics_2014, akita_Experimental_2017, schenz_Mathematical_2017}. 
Both morphodynamic remodeling and the migration velocity are closely tied to the self-organized patterns of contraction~\cite{kakiuchi_Multiple_2006, zhang_Symmetry_2019, fleig_Emergence_2022}, drawn from a large set of contraction modes connected to behavioral adaptation set by the environment~\cite{fleig_Emergence_2022, wilkinson_Flow_2023, boussard_Adaptive_2021}. 
Despite these insights, the underlying mechanics and how these mechanics are adapted during the decision-making process remain largely unknown.\\

\begin{figure*}[ht]
    \centering
    \includegraphics[width=17.4cm]{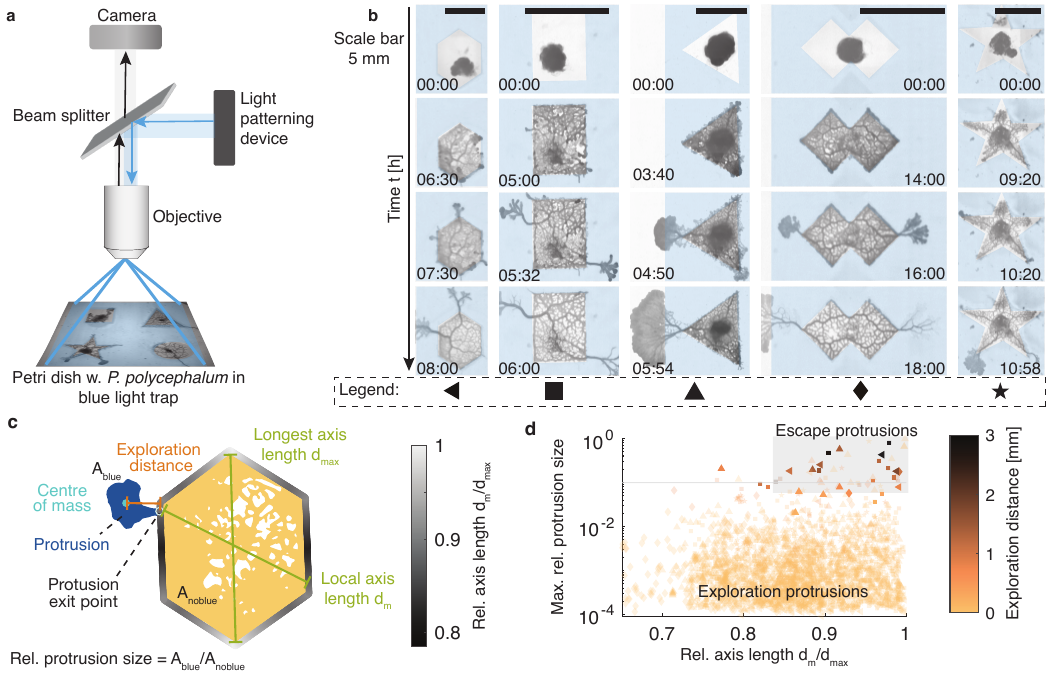}
    \caption{Regardless of the shape of the blue light trap, \textit{P.~polycephalum} escapes along the longest axis of the shapes. (a) Bright-field microscopy set up with blue light (470~nm) illuminated shapes as traps for \textit{P.~polycephalum} plasmodia exploring their environment with protrusions. (b) Exemplary \textit{P.~polycephalum} exploration in different shapes results in escape along longest axis. (c) Illustration of parameters to quantify exploration via protrusion size $A_\mathrm{blue}$, exploration distance, and exit point along the trap boundary. The trap boundary is parameterized by the relation of local axis length $d_\mathrm{m}$ at the exit point normalized by the longest axis length $d_\mathrm{max}$ of the trap. (d) Small protrusions emerge all around the trap boundary (exploration protrusions), yet escapes only happen close to the longest axis within the shape (escape protrusions highlighted by gray box and 100\% opacity). Exploration distance: Euclidean distance of protrusion center of mass to protrusion exit point on trap boundary.}
    \label{fig:fig1}
\end{figure*}

\noindent In our study, we utilized blue light confinement to entrap \textit{Physarum polycephalum} into polygonal shapes and closely monitor its exploratory dynamics leading up to escape. We found that \textit{P.~polycephalum} consistently escapes by forming a protrusion aligned with the longest axis of the confining geometry, which coincides with the dominant mode of its peristaltic wave. During exploration, however, the organism transiently forms protrusions in other directions along the trap boundary. Principal component analysis of the contraction dynamics reveals that these exploratory behaviors reflect dynamic switching between distinct contraction modes. The trap shape ultimately sets the mode most efficient for transport, allowing pressure to build up along the longest axis and driving the plasmodial escape. Our results suggest that decision-making in \textit{P.~polycephalum} involves active mechanical processes, reflecting out-of-equilibrium dynamics where the organism transiently adopts less optimal transport states during unconfined exploration. These findings offer new insight into how behavior and decision-making emerge in organisms lacking a central nervous system.

\section*{Results}
\subsection*{Exploration events due to localized protrusions lead to escape along longest axis}
\noindent We use \textit{Physarum polycephalum}'s strong avoidance reaction to blue light ~\cite{bauerle_Living_2020, chen_Network_2023, nakagaki_Action_1996, bialczyk_Action_1979,mori_Spatiotemporal_1986} to trap its plasmodial network in planar geometric shapes and follow individual exploration until escape from the trap. Polygonal shapes at a wavelength of 470~nm are generated by a light patterning device and passed through a beam splitter on top of Phytagel plates of reduced autofluorescence~(Fig.~\ref{fig:fig1}a). Traps have the shape of a hexagon, square, triangle, zigzag diamond polygon, and a star with an similar areal size between 10-30~mm$^2$ to ensure that observed behavior is independent of the trap size. 
\textit{P.~polycephalum} networks are initially grown on plain nutrient-free agar before being transferred as small blobs directly into the traps projected onto Phytagel plates. This starving conditioning forces the specimen to migrate and escape from the traps. Note that no nutrients are present in the experimental setup, so a plasmodium cannot forage successfully.
We take time series of bright field images of 2-4 plasmodia simultaneously at 2-minute intervals over 24~h (Fig~\ref{fig:fig1}b, Supplemental Movies~1-5) to observe how plasmodia explore their surroundings until they escape. Additionally, we take data sets with a higher frame rate of 3~s or 6~s between images to capture the individual specimen's continuous reorganization and contraction dynamics.
\begin{figure*}
    \centering
    \includegraphics{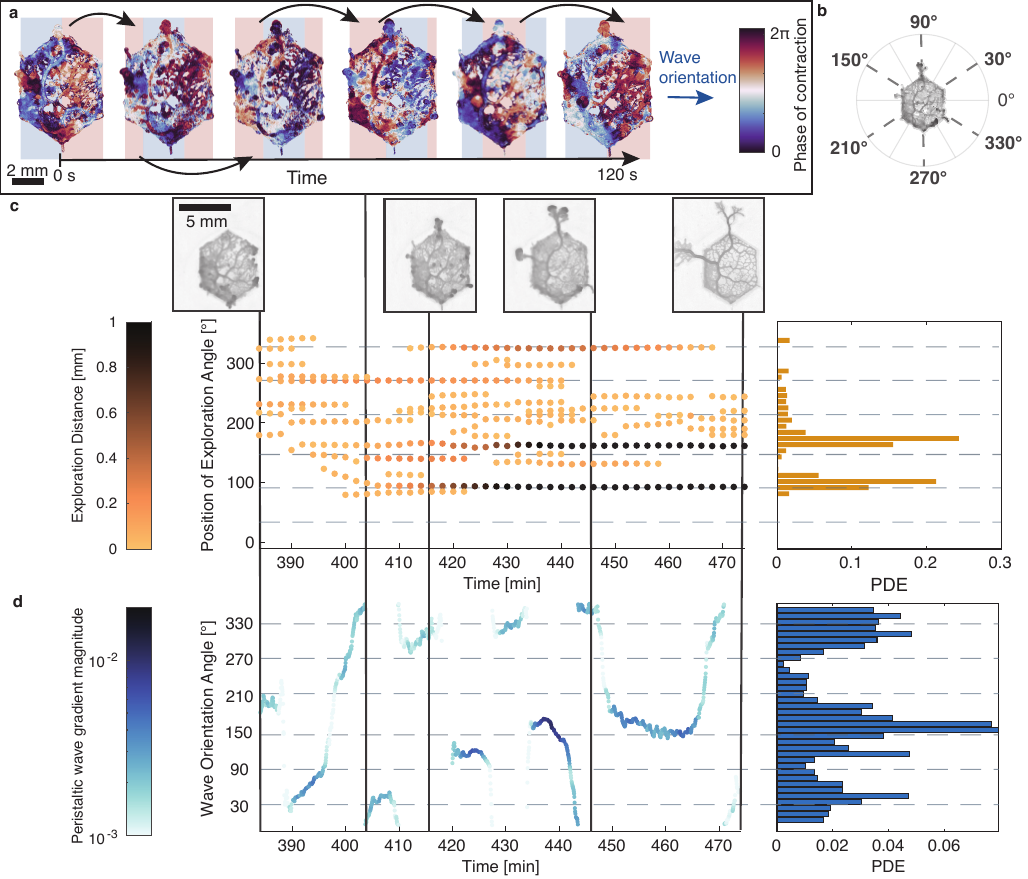}
    \caption{During exploration angular protrusion location aligns with the orientation of peristaltic wave. (a) Extraction of peristaltic wave direction (blue arrow) from sheet-like plasmodial structures throughout 120~s-periods is perpendicular to the longest axis in this example. The underlying blue and red boxes serve as visual guides, indicating regions with a visually averaged phase in the ranges [0, $\pi$] (blue) and [$\pi$, $2\pi$] (red). A black arrow highlights the movement of the reddish phase region. (b) Orientation of longest axes (gray dashed lines) within hexagonal shaped trap. (c) During the exploration phase, the angular component of the center of mass of localized protrusions over time is defined as the trap being completely filled and lasting until escape. Protrusions cluster around the longest axes of the trap but emerge elsewhere as well. Exploration distance is measured using a minimal Euclidean distance of the protrusion center to trap the boundary. (d) Orientation of peristaltic wave over time mainly follows longest axes of shape with gradient magnitude being especially large and wave more stable along the escape axis of $~150^\circ$. Histograms in (c) and (d) display the relative probability (PDE) of the event count. The gray dashed lines indicate the longest axes within the hexagonal shaped trap. (See Supplemental Movies~1-5)}
    \label{fig:fig2}
\end{figure*}
\textit{P.~polycephalum} starts growing within 60~min after inoculation under the microscope. The individual plasmodia first nearly fill out their traps with a relatively dense network; see Fig~\ref{fig:fig1}b. Then, the plasmodia explore their surroundings by forming small localized protrusions extending into the blue light. Eventually, they decide on an escape route over which large parts of their body mass are transported over the blue light area out of the trap. We stop our quantitative analysis of the exploration behavior once the escaping protrusion touches the outside boundary of the blue light trap, as we here focus on the exploration before the plasmodia finalize their decision on the escape route.\\

\noindent Protrusions allow \textit{P.~polycephalum} to gather information about the environment at the periphery of the network~\cite{dussutour_Amoeboid_2010}. We therefore quantify protrusion dynamics into the blue light by three parameters: their size, exploration distance, and exit point along the trap boundary, see Fig.~\ref{fig:fig1}c.
In detail, we determine the maximum size of each individual protrusion extending into the blue light, denoted as $A_{\text{blue}}$, and normalize it by the total network area within the trap, $A_\text{noblue}$. This ratio provides a measure of the relative protrusion expansion. 
Furthermore, we analyze the protrusion’s movement within the blue light by calculating its exploration distance, defined as the minimal Euclidean distance between the protrusion’s center of mass and the trap boundary. This metric captures how far the protrusion extends into the blue light.
Finally, we identify the protrusion exit point as the moment when its center of mass first crosses the trap’s boundary. To parameterize this exit location, we introduce the relative axis length $d_\mathrm{m}/d_\text{max}$. Here, $d_\mathrm{m}$ represents the longest possible internal path through the shape at the exit point, normalized by $d_\mathrm{max}$, the longest axis of the entire shape. This ratio quantifies how closely the protrusion’s escape aligns with the longest axis of the shape (see Supplement Fig.~S1 for all shapes~\cite{schick_supp_2026}).\\

\noindent Mapping protrusion size, exploration distance, and their exit point, parametrized by relative axis length, across 32 plasmodia in five different shapes reveals a clear pattern in \textit{P.~polycephalum}'s exploratory and escape behavior. Independently of the trap shape, protrusions emerge along the entire trap boundary but remain small in size $A_\mathrm{blue}/A_\mathrm{noblue}  \lesssim 0.1$ and extend only short distances into the blue light (light orange shaded data points in Fig.~\ref{fig:fig1}d). However, the few protrusions that ultimately escape by crossing the outer trap boundary extend the farthest (exploration distance $>$ 1~mm), are consistently larger, and tend to exit near trap corners, marking the longest axis within the trap shapes with $d_\mathrm{m}/d_\mathrm{max}\rightarrow 1$ (highlighted with 100\% opacity, Fig.~\ref{fig:fig1}d). 
While these escape positions often coincide with regions of higher curvature of the trap boundary, escape alignment with the longest axis persists even in nearly circular traps with strongly reduced curvature gradients (Supplement Fig.~S5~\cite{schick_supp_2026}), as well as in the zigzag diamond polygon with corners with comparable curvature but distinct axis lengths (Fig.~\ref{fig:fig1}b shape~4). This suggests that the longest axis represents the optimal escape route, consistent with observations that contraction patterns tend to align along the same axis.\\

\noindent The escape along the longest axes aligns with previous observations that \textit{P.~polycephalum}'s peristaltic wave self-organizes along the longest structural axes of its network~\cite{alim_Random_2013}. However, here, this alignment emerges only at the onset of escape. In contrast, over the course of the 1.5~hours long exploration phase leading up to escape, protrusions form in multiple directions without a dominant orientation. 
This shift from multidirectional exploration to a dominant escape trajectory raises the question how internal dynamics reorganize to support this transition. In particular, how does \textit{P.~polycephalum} explore multiple directions despite the peristaltic wave’s known tendency to align along the longest axis?


\subsection*{Orientation of peristaltic wave aligns with location of protrusion}
\noindent To test the direction of the peristaltic wave during the exploration phase from its onset at complete filling of the traps to its end with escape, we extract the contraction phase of every pixel within a plasmodial network from the bright-field intensity changes corresponding to tubular contractions in high time resolution data sets with 6~s between frames~\cite{takagi_Emergence_2008, takagi_Annihilation_2010, alonso_Oscillations_2016, zhang_Selforganized_2017} (Fig.~\ref{fig:fig2}a, Supplemental Movie~2). We quantify the wave direction by calculating the gradient of the smoothed phase pattern and averaging the dominating direction over one contraction period of 120~s~\cite{kamiya_Physical_1981, wohlfarth-bottermann_Oscillatory_1979}. Here, dominant refers to the prevailing contraction direction across the network, capturing the main axis of peristaltic wave propagation despite local deviations.
Following the exemplary contraction phase over a single contraction period, we observe the dominant wave to be oriented perpendicular to the longest axis, see visual guide in Fig.~\ref{fig:fig2}a. 
Now, to compare wave direction with the center of mass of the protrusions, we project both onto their angular orientation from 0$^\circ$ to 360$^\circ$ relative to the trap center (Fig.~\ref{fig:fig2}b) for an individual plasmodia in the hexagonal trap. While the wave direction is extracted from high time resolution data, protrusion growth occurs at a slower timescale and is, therefore, analyzed in a subset of the individual datasets at 2~min intervals.

\begin{figure}[ht]
    \centering
    \includegraphics[width=8.6cm]{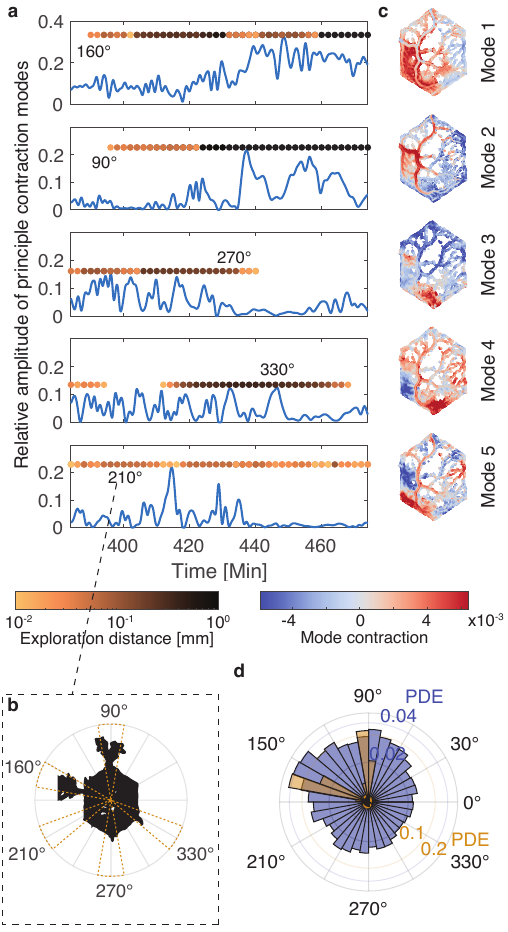}
    \caption{Active switching of contraction modes enable different exploration positions. (a) Timelines of relative amplitude of principle contraction modes 1-5 with corresponding protrusion's exploration distance. (b) Global mask of protrusion growth with windows chosen to merge localized protrusion exploration (orange highlight). (c) Structure of the five highest-ranking PCA modes in the hexagonal data set. (d) Histogram of gradient orientation in PCA modes weighted by average relative amplitude (blue) and the corresponding exploration frequencies (orange).}
    \label{fig:fig3}
\end{figure}
\noindent At the onset of exploring the trap's surroundings, the plasmodium covers the entire blue light-free region inside the trap, extending and retracting protrusions in multiple directions. Over approximately 1.5~hours leading up to escape, protrusions briefly push into the blue light before retracting. This behavior is seen here in a hexagonal trap (Fig.~\ref{fig:fig2}c), and similar patterns can be observed in other trap geometries (see Supplement Figs.~S2-S4~\cite{schick_supp_2026}).
At first glance, protrusion growth appears directionally unstructured. Protrusions extend in many directions but rarely persist, often retracting after limited growth (light orange data points, Fig.~\ref{fig:fig2}c). However, a closer look reveals a bias: most protrusions cluster along the longest axes of the hexagon at [$30^\circ$, $90^\circ$, $150^\circ$, $210^\circ$, $270^\circ$, $330^\circ$] (highlighted by dotted lines in Fig.~\ref{fig:fig2}c). Not all of these escape attempts are equal. While smaller, short-lived protrusions appear at $330^\circ$ and between $210^\circ$–$270^\circ$ (predominantly on the left side of the trap), the most pronounced protrusions grow at $90^\circ$ and $160^\circ$ (exploration distances significantly above 1~mm). Yet, even among these protrusions, a distinction emerges: the protrusion at $90^\circ$ initially pushes into the blue light but ultimately retracts, whereas the protrusion at $160^\circ$ persists, ultimately defining the escape route (see histogram with relative probability of exploration events, Fig.~\ref{fig:fig2}c).\\

\noindent We compare protrusion exploration with peristaltic wave orientation to understand why the protrusion at $160^\circ$ persists while others retract. The peristaltic wave takes time to settle into the longest-axis orientation. During the onset of smaller protrusions, the peristaltic wave continuously reorients across the network (Fig.~\ref{fig:fig2}d). However, even in this phase, the wave primarily fluctuates between the different long axes of the trap, switching between $30^\circ$, $150^\circ$, and its opposite $330^\circ$.
Despite this, a large protrusion initially emerges at $90^\circ$, appearing predominantly inconsistent with a dip in the dominant wave orientation. As the exploration phase progresses, a shift occurs around 20~min before the escape: rather than fluctuating between different directions, the peristaltic wave stabilizes around $150^\circ$ (most significant gradient magnitude, Fig.~\ref{fig:fig2}d). As this direction is the ultimate escape route, a sustained progression of the peristaltic wave appears to be required for effective mass transport for the escape. These observations are generalizable to the other trap shapes (Supplement Figs.~S2-S4~\cite{schick_supp_2026}).\\

\noindent While protrusion growth is aligned mainly with peristalsis-driven mass reallocation, the emergence of the pronounced protrusion at $90^\circ$, despite the peristaltic wave predominantly propagating toward $30^\circ$, indicates that protrusion initiation cannot be explained solely by the instantaneous dominant wave direction. This mismatch highlights a limitation of dominant-wave analysis: it emphasizes the most persistent global propagation direction but does not capture structured contraction modes that are repeatedly expressed and contribute substantially to exploratory behavior. The peristaltic wave is highly dynamic during exploration, fluctuating between different long-axis orientations before ultimately stabilizing at $150^\circ$ prior to escape. Since stable peristalsis appears necessary for sustained mass transport and successful escape, the critical question becomes: What determines the selection of the peristaltic wave direction? Resolving this question is essential to understanding how \textit{Physarum polycephalum} transitions from broad exploration to a directed escape response.

\subsection*{Varying activation of contraction modes during exploration}
\noindent To study the onset of directionality in the peristaltic wave, we turn to a detailed analysis of the contraction patterns in \textit{P.~polycephalum} with Principal Component Analysis (PCA). We apply PCA onto high time resolution data of the exploration phase to decompose contraction patterns, i.e.~the intensity changes across all pixel present throughout the entire exploration phase, into a set of $N$ orthonormal eigenmodes, $k\in N$ of time-varying eigenvalue amplitude ${a}_{k,i}$~\cite{fleig_Emergence_2022}. 
We define the activity $\Pi_{k,i}$ of each mode $k$ at time point $i$ as the relative amplitude $ \Pi_{k,i} = \frac{\Tilde{a}_{k,i}}{\sum_\nu\Tilde{a}_{\nu, k,i}}$, where $\Tilde{a}_{\nu, k,i}$ is the squared amplitude of the mode coefficient $a_{k,i}$ normalized by the envelope of squared coefficients across all modes $\nu$ at that time point~\cite{fleig_Emergence_2022}.
Mapping the activity of the five highest-ranking contraction modes over time (Fig.~\ref{fig:fig3}a) reveals significant temporal activity variation during the exploration phase. We compare mode activity with the exploration distance of protrusions forming at key angular positions to investigate whether these dynamic patterns relate to directional growth behavior. Specifically, we focus on five dominant directions [$90^\circ$, $160^\circ$, $210^\circ$, $270^\circ$, $330^\circ$], which correspond to the primary regions of protrusion growth identified in previous analyses. These directions largely align with the corners of the hexagonal trap and, therefore, with the longest internal axes (highlighted in orange in Fig.~\ref{fig:fig3}b). 
Protrusion growth events are binned according to their angular position within these five dominant directions. The corresponding exploration distances are plotted in logarithmic space to emphasize both the onset and end of the growth of protrusions independently of size.
At the start of the exploration phase, \textit{P.~polycephalum}'s exploratory behavior is driven by a set of distinct contraction modes (modes~2–5), whereas the transition to escape behavior is primarily associated with mode~1 (Fig.~\ref{fig:fig3}a). This reinforces our understanding that different phases of behavior, exploration versus escape, are supported by distinct underlying contraction dynamics. In other words, specific principal contraction modes are selectively activated to enable different types of protrusion growth.\\

\noindent PCA modes represent standing waves, with their superposition giving rise to peristaltic motion and intracellular flow~\cite{fleig_Emergence_2022}. Therefore, to examine how contraction dynamics are spatially organized, we focus on the orientation of the five highest-ranking PCA modes (Fig.~\ref{fig:fig3}c). We observe that mode~1 displays a clear standing wave along the left-right axis, aligning with the dominant escape direction. In contrast, modes~2 and~3 exhibit wave orientations more aligned with the top–down axis, while the patterns in modes~4 and~5 are more complex and harder to interpret visually. To systematically quantify these orientations, we use surface fitting and compute the corresponding gradient fields to extract the orientation angle of the PCA mode. Directional biases were captured using normalized histograms, with each mode’s contribution weighted by its average activity over time. The final cumulative histogram reflects the dominant gradient directions across all contraction modes within the confined plasmodium while accounting for the varying significance of individual modes.
The resulting orientation distribution, shown in Fig.~\ref{fig:fig3}d, reveals that the combined activity of multiple modes drives exploration biased towards the longest axes of the hexagonal shape.\\

\noindent These observations raise the question: What determines which contraction mode dominates at a given time? Since contractions in \textit{Physarum polycephalum} are closely linked to fluid flow, it is plausible that internal fluid mechanical constraints shape the temporal evolution of mode activity. Rather than being random, using different modes might reflect an underlying strategy to organize transport, supporting the transition from exploration to committed escape.

\subsection*{Confinement sets contraction mode optimized for transport}
\begin{figure}[ht]
    \centering
    \includegraphics[width=8.6cm]{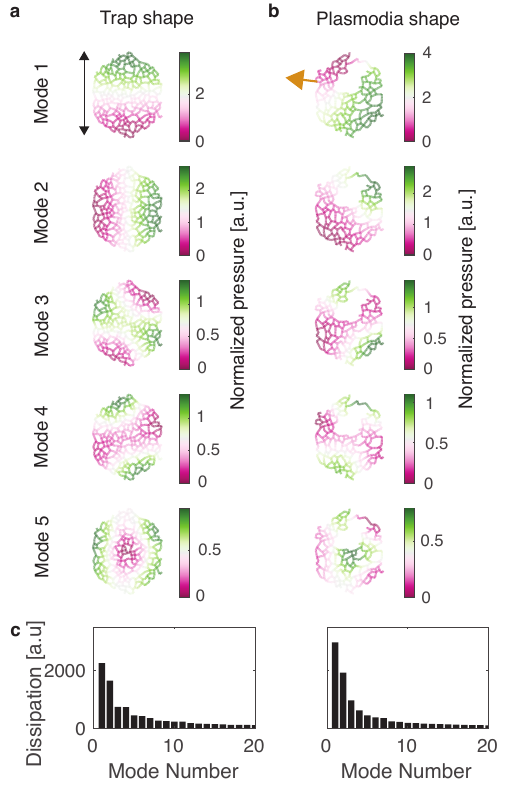}
    \caption{Confinement into shape enforces most efficient transportation pattern along escape route demonstrated on theoretical network spanning the polygonal trap shapes. (a) Top 5 ranking theoretical pressure modes acquired from Singular Value Decomposition on efficiency for highest transport with first modes accounting for most of the dissipated energy (c). (b) Reorientation of pressure modes driven by network topology aligns first mode with escape of \textit{P.~polycephalum} indicated in orange arrow.}
    \label{fig:fig4}
\end{figure}
\noindent The contractions in \textit{P.~polycephalum} are closely linked to fluid flow throughout the network. 
A large body of theoretical work on optimal flow networks~\cite{murray_Physiological_1926, katifori_Damage_2010, corson_Fluctuations_2010, ronellenfitsch_Phenotypes_2019, hu_Adaptation_2013}, including those operating far from equilibrium~\cite{akita_Experimental_2017}, suggests that such systems tend to minimize energetic costs associated with transport. 
Taking the non-illuminated trap shapes, we generate a theoretical network of nodes and edges by randomly distributing nodes across the polygonal shape, subject to minimum inter-point distances that enforce an approximately uniform node density, following the work of Wilkinson et al.~\cite{wilkinson_Flow_2023}. As contraction modes are mostly independent of the tube geometry~\cite{wilkinson_Flow_2023}, a uniform network is a good approximation for a dense plasmodium.

Each tube connecting nodes $(ij)$ has a time dependent volume $V_{ij}(t) = \pi l a_{ij}(t)^2$ with constant length $l$ and in time varying radius $a_{ij}(t)$. Transport costs in the flow network are given by the dissipation which for a single tube can be written as $D_{ij} = \frac{Q_{ij}^2}{\kappa_{ij}} + \frac{1}{12\kappa_{ij} } \left(\frac{\partial V_{ij}}{\partial t}\right)^2$ with the volume flow rate $Q_{ij}$ and hydraulic conductance $\kappa_{ij} = \frac{\pi a_{ij}^4}{8\mu l}$ with fluid viscosity $\mu$~\cite{murray_Physiological_1926}. 
The first part of the equation is the Hagen-Poiseuille dissipation, quantifying the cost of net fluid transport. The second part accounts for dissipation of the bidirectional flow arising from rhythmic tubular contractions. This part reflects the energetic cost of oscillatory dynamics that do not contribute directly to net displacement but are essential for internal mass redistribution. 
The total dissipation of the network can then be calculated by summation over all individual tubes. 
Wilkinson et al.~\cite{wilkinson_Flow_2023} found a fundamental relationship connecting the volumetric flow rate with the volume change for the individual tubes of the full network $\vec{Q} = \bm{\Gamma} \frac{\partial \vec{V}}{\partial t}$ where the linear mapping $\bm{\Gamma}$ incorporates both the network topology and the geometry of the tubes. 
This leads to a matrix representation of the dissipation $D_{\mathrm{tot}} = \frac{\partial \vec{V}}{\partial t}^\top \bm{\Gamma}^\top \bm{\kappa}^{-1}\bm{\Gamma} \frac{\partial \vec{V}}{\partial t} + \frac{\partial \vec{V}}{\partial t}^\top \bm{\kappa}^{-1} \frac{\partial \vec{V}}{\partial t} $ in terms of tube's volume change, where $\bm{\kappa}$ is a diagonal matrix with each edge's conductances. \\

\noindent Energy should primarily be spent to generate forward-directed flow for maximal transport efficiency. Any reverse or oscillatory flow that does not contribute to net transport constitutes a loss. We, therefore, define transport efficiency as the ratio of the dissipation of  forward-directed Hagen-Poiseuille flow to the dissipation `wasted' for bidirectional flow.
To identify the volume change patterns that lead to highest transport efficiency, we apply Single Value Decomposition (SVD) to the matrix form of the dissipation ratio, decomposing the transport efficiency into orthogonal volume change pattern (See Supplement Section~III~\cite{schick_supp_2026}). The resulting eigenvectors correspond to specific volume change patterns, while the corresponding eigenvalues represent the transport efficiency associated with each pattern. 

\noindent From the volume change eigenvectors, we directly calculate the corresponding flow patterns, as well as pressure modes, where each pressure mode is averaged across the two ends of the tube and shifted such that the minimum pressure in the network is set to zero (Fig.~\ref{fig:fig4}a).
The first few modes of the decomposition account for the majority of the dissipation invested in directed flow (Fig.~\ref{fig:fig4}c), indicating that they carry the highest fluid transport across the network.
In the hexagonal trap, the most efficient mode facilitates transport along the vertical (top-down) axis, while the second mode supports transport along the horizontal (left-right) axis (Fig.~\ref{fig:fig4}a). While an ideal hexagon would display symmetric modes, a slight vertical elongation of the experimental network, arising from optical aberrations, lifts this degeneracy, with the corresponding symmetry partners manifesting only at higher modes (modes~3 and~4). The vertical orientations of the first two modes are consistent with the principal contraction patterns extracted from the experimental data via PCA (Fig.~\ref{fig:fig3}c). 
However, when considering the actual topology observed in the experiment, specifically the large hole in the top-right part of the network, we find that the theoretically optimal transport direction shifts. By incorporating this hole into the theoretical model, the dominant transport mode rotates to align along the $120^\circ$-axis (Fig.~\ref{fig:fig4}b).
Together with the second mode, oriented along the $30^\circ$-axis, this results in the most efficient transport aligned with the observed escape direction, approximately $160^\circ$. This demonstrates how local network topology can significantly reshape global transport modes. These transport patterns are not static. Because the contraction dynamics form standing waves, flow and pressure directions periodically reverse throughout a contraction cycle. Still, the strongest average gradients remain aligned with the dominant transport modes.

\noindent We conclude that the confinement into a blue light free trap enforces \textit{P.~polycephalum} to escape along the most efficient transportation phase pattern. The highest pressure gradients arise along these theoretical pathways, likely driving the large mass reallocation required to escape the trap.
Nonetheless, the exploration phase of 1.5~h in this example shows that the system does not solely follow the most efficient route but rather samples a wide range of less efficient protrusion growth options first. These repeated explorations suggest that the system does not exclusively follow the most optimal transport direction. Instead, it dynamically tests multiple configurations before settling into the escape trajectory. In comparison, freely migrating plasmodia of the same size can reach migration velocities of around 2~cm/h~\cite{kuroda_Allometry_2015}, suggesting
that even a less optimal escape path might suffice under no confinement. This behavior reveals the signature of an out-of-equilibrium system that balances exploratory flexibility with mechanical efficiency in its decision-making process.

\section*{Discussion}
\noindent Migration decisions in \textit{P.~polycephalum} are shaped by environmental and energetic constraints. When \textit{P.~polycephalum} is confined in a trap under blue light, we observe that initial exploratory protrusions extend in all directions before the organism consistently escapes along the longest axis of the trap. The peristaltic wave responsible for fluid transport and migration rapidly switches during the exploratory phase but stabilizes along the trap axis at the point of escape.
Further analysis of contraction patterns reveals that different contraction modes activate individual exploration protrusions. Ultimately, the plasmodium settles into the contraction mode, enabling the most efficient transport determined by the confinement geometry to escape the trap. These results suggest that decision-making in extended transport networks arises from a dynamic reorganization of internal mechanics shaped by environmental constraints.\\

\noindent Research on \textit{P.~polycephalum} has primarily focused on the outcomes of decision-making. The plasmodium is known to make smart foraging decisions~\cite{latty_Food_2009, nakagaki_MinimumRisk_2007, tero_Rules_2010, schick_Dynamic_2024} and has demonstrated remarkable problem-solving capabilities in various path finding tasks~\cite{nakagaki_Smart_2001, tero_Physarum_2006, reid_Decisionmaking_2016}. Our work advances this perspective by directly linking \textit{P.~polycephalum}’s internal contraction dynamics to its emergent behavior. Rather than assuming the organism always selects the optimal solution, we show that decisions are driven by internal fluid transport and shaped by environmental constraints. While transport networks in biology are often studied through the lens of optimization~\cite{murray_Physiological_1926, katifori_Damage_2010, corson_Fluctuations_2010, ronellenfitsch_Phenotypes_2019, hu_Adaptation_2013}, our findings suggest that \textit{P.~polycephalum} relies on a more flexible strategy. Freely migrating plasmodia exhibit switching between contraction modes~\cite{fleig_Emergence_2022, wilkinson_Flow_2023}, and without geometric constraints, may explore less efficient configurations before settling into one that supports effective transport. This behavior reflects an active, out-of-equilibrium process in which physical dynamics drive decision outcomes.\\

\noindent Contraction patterns in \textit{P.~polycephalum} are influenced by chemical pathways such as calcium signaling~\cite{kscheschinski_Calcium_2024} and blue light-induced mechanical changes to the actin cortex~\cite{chen_Network_2023, bauerle_Living_2020}. Blue light has also been shown to alter metabolism and cell differentiation~\cite{schreckenbach_Bluelight_1981}, potentially affecting broader physiological responses. In partially illuminated specimens, such combined effects may contribute to memory-like behavior, consistent with previous suggestions that morphological memory arises from localized softening within the network~\cite{kramar_Encoding_2021} and learning arising from biochemical pathways~\cite{eckert_Biochemically_2024}. However, our findings suggest that while mechanical changes may occur throughout the entire network during decision-making, local softening alone cannot account for the observed escape behavior. 
Instead, the network architecture itself appears to play a central integrative role by constraining the set of available contraction modes that govern transport. The relative amplitude and persistence of these modes are governed by environmental stimuli such as blue light~\cite{fleig_Emergence_2022}. Our analysis now reveals a change in relative amplitude of principal contraction modes following exploratory protrusions extending into the blue light. The continuous formation of exploration protrusion activated by specific contraction modes therefore seems to drive a selection among the contraction modes which favor the contraction mode with most efficient transport to prevail for the longest duration to form a sufficiently large protrusion for escape. The observed adaptive behavior is therefore a result of the dynamic interplay between network architecture, environmental stimulation, and contraction mode selection.

\noindent All in all, our findings reveal how decision-making can emerge from physical dynamics in a non-neuronal, network-forming organism. By linking \textit{P.~polycephalum}’s behavioral transitions to internal contraction patterns and environmental constraints, we demonstrate that adaptive behavior can arise from the self-organization of internal flows rather than centralized computation. This perspective not only advances our understanding of decision-making in non-neuronal systems but also contributes to a broader reevaluation of what constitutes a decision in living matter. By grounding these processes in measurable mechanical principles, \textit{P.~polycephalum} becomes more than a biological curiosity, it offers a powerful model for rethinking the foundations of decision science, behavior, and cognition from the bottom up.\\

\section*{Materials and Methods}
\subsection*{Culturing and imaging of \textit{P. polycephalum}}
\noindent The microplasmodia of \textit{P.~polycephalum} (Carolina Biological Supplies) are cultivated in a liquid culture at 25$^{\circ}$C. Plasmodia are transferred into two to three 0.5x0.5~cm wells in a 1.5\% (w/v) plain agar plate. The agar plate is incubated for at least 20~h at 25$^{\circ}$C in the dark, such that large plasmodial networks are formed before translation of small volumes of scraped plasmodia to a Phytagel plate subject to blue light traps within the microscope for imaging at reduced autofluorescence of the Phytagel plate.  

\noindent Imaging is conducted with a Zeiss Axio Zoom V.16 equipped with a Hamamatsu ORCA-Flash 4.0 digital camera and a Zeiss PlanNeoFluar 0.5×/0.125 objective, providing a spatial resolution of 50-90~px/mm. 
A green filter (740 Aurora Borealis Green, LEE Filters) is added above the light source to block ambient stimuli light~\cite{bialczyk_Action_1979}. 
During the experiment, the digital mirroring device (DMD) \textit{Mightex Polygon} continuously projects pixelated blue light patterns with an LED wavelength of 470~nm onto a Phytagel plate. The light intensity of the LED is downregulated to 1-2\% such that the plasmodia can still cross over the light trap. 
Bright-field images are acquired at least every 5~min for about 18~h for experiments that track the escape directions of the plasmodia. Additional experiments for quantification of contractions recorded frames at 6~s intervals within acquisition blocks of 1~h each.

\subsection*{Exploration analysis}
\noindent Bright-field images are stored in the Carl-Zeiss-Image datatype and converted into 8-bit tiff files with a custom-built videotool based on the Python library \textit{czifile}. 
A rolling-ball algorithm removed uneven illumination in the background on each tile. \textit{P.~polycephalum} networks were extracted with a custom-written MATLAB (The MathWorks) code, creating a binary image by adaptive intensity thresholding based on fuzzy c-means cluster~\cite{semechko_Fast_2023} and closing single pixels. 
Using the MATLAB built-in functions \textit{bwconncomp}, first, the individual plasmodial networks in each frame were detected and assigned to the plasmodia of the previous time steps, enabling a continuous time series of pixel lists for the individual specimen. 
In a second step, morphological features of the individual plasmodia like network area $A$, convex area size $A_c$, bounding box, centroid, and perimeter were extracted over time with the MATLAB function \textit{regionprops}. 
Third, for each plasmodium, the protrusions growing into the blue light are extracted by combining the plasmodium's pixel list with the illuminated region. The blue light illuminated regions are extracted from the shadow of the non-illuminated area with the \textit{drawrectangle/ drawpolygone or drawellipse} tool in MATLAB. MATLAB's function \textit{regionprops} is employed to extract protrusions' morphological features like area and centroid.

\subsection*{Phase analysis}
\noindent The continuous time series of pixel lists for individual plasmodia and its associated intensities is used to analyze the peristaltic contraction phase. Following the approach in~\cite{chen_Network_2023}, we first detrend the individual pixel intensity time series with a moving average filter to remove the signal oscillations on long time scales and smooth the resulting pixel intensity time series with a Gaussian kernel. 
We then use a Hilbert transform to extract the instantaneous phase of the intensity time series for all pixels of a plasmodium, following~\cite{bauerle_Spatial_2017}. The phase values are rescaled to the interval~$[0,2\pi]$. 
For the directional analysis of the phase wave within the plasmodium, we first create a surface fit from the scattered phase values at every time frame using the MATLAB \textit{gridfit} function~\cite{derrico_Surface_2023}. 70 grid points in the x and y directions combined with a smoothness parameter of 30 are used for the phase surface fitting. 
Due to this phase fitting step, the directional analysis is limited to the dominant phase wave pattern on the spatial scale of the plasmodium.
Considering the surface fit grid points within the respective blue light shape for every time-series frame, we obtain the two-dimensional gradient field. 
We further approximate the temporal change of the phase at every grid point via finite differences (first order at the first and last frame of the time series, second order otherwise) between frames in the time series. 
We then combine the spatial and temporal information to estimate the direction of the traveling wave by multiplying the spatial phase gradient at every considered grid point with the negative temporal phase change into phase wave information, respectively. 
To render robust results, we combine the phase wave information from several frames within a moving window (moving stride of one frame), which corresponds to a typical contraction period of \textit{P.~polycephalum} of~120~s\cite{wohlfarth-bottermann_Oscillatory_1979, kamiya_Physical_1981}. 
For every moving window, we compute the average temporal phase change-weighted gradient vectors at the grid points and summed the resulting vectors to obtain a total gradient vector. 
Employing the four-quadrant inverse tangent MATLAB function on the x and y components of this total gradient vector, we estimate the wave direction of the phase in a~$[0°,360°]$ range in the two-dimensional plane of the plasmodium. 
In addition, we calculate the magnitude of the total gradient vector, normalized by the number of considered grid points within the shape, to be able to assess the uncertainty of the direction estimate within a certain time window. 

\subsection*{Principal component analysis}
\noindent We perform a Principal Component Analysis (PCA) to analyze the continuous time series of pixel intensities associated with individual plasmodial contractions. For this analysis, we consider only pixels that remained present throughout the entire time window of 90~min before the plasmodium escapes its trap. The bright-field intensity values of each pixel are detrended using a moving average to remove baseline trends of 120~s contractions. Here, the pixels serve as features, while the bright-field intensities are treated as PCA observations, processed by MATLAB's \textit{pca}~function. To ensure consistency, the resulting modes are normalized using \textit{normc} prior to calculating the relative mode amplitudes to acquire the activity of the individual contraction modes.

\subsection*{Transport efficient flow modes}
\noindent We generated theoretical networks of non-illuminated trap shapes to identify contraction modes optimized for transport efficiency. The theoretical network was created within the binary mask of the polygonal trap shapes using constrained random sampling, iteratively placing points at random locations subject to a minimum inter-point distance, enforcing an approximately uniform density within the polygons. The networks are modeled by an oriented graph $G$ with $m$ labeled nodes, created as equally spaced grid points within the polygonal shapes, and $n$ edges, corresponding to tube elements with uniform length and thickness, thereby eliminating any inherent network hierarchy. The edges are calculated by the MATLAB function \textit{RelativeNeighborhoodGraph}, based on the Euclidean norm. 
The resulting graph is further refined by discretizing the structure and removing any loops between two nodes. 
This processed graph serves as the foundation for identifying volume change patterns in the individual tubes, thereby optimizing transport efficiency, by applying Singular Value Decomposition (SVD) to the matrix form of the ratio of the individual parts of the dissipation relation, see Supplement Section III~\cite{schick_supp_2026}. The computed volume change patterns are ranked by their efficiency for directed transport, enabling volume flow and pressure map calculations on the theoretical networks. 

\subsection*{Mode directionality analysis}
\noindent For the directional analysis of both the PCA and pressure modes within the plasmodium, we again initially use the MATLAB gridfit function~\cite{derrico_Surface_2023} for surface fitting of 70 grid points in the x and y direction combined with a smoothness parameter of 10.
Considering the surface fit grid points within the respective blue light shape, we obtain the two-dimensional gradient field for every mode and calculated the angles of all vectors.
We generate normalized histograms for every mode with $10^\circ$ bin widths in the range~$[0°,360°]$. 
The individual mode angle bin counts are weighted by the average mode activity before normalizing again for the final cumulative histogram, summarizing the dominant gradient directions of the individual modes within the confined plasmodium while considering the mode's significance.

\section*{Data Availability}
The collected experimental data and code for simulation is available in the mediaTUM repository~\cite{schick_mediaTUM_2026}.

\begin{acknowledgements}
\noindent This work received funding from the European Research Council (ERC) under the European Union’s Horizon 2020 research and innovation program (grant agreement No. 947630, FlowMem) to K.A., the Deutsche Forschungsgemeinschaft DFG through INST 95/1634-1 FUGG to K.A. and the Human Frontier Science Program Organization through Research Grant number RGP0001/2021 to K.A. and M.R.. We thank Arne Rosenberg for setting up the Mightex Polygon and for providing data that helped shape the research question. 
\end{acknowledgements}
\section*{Author Contributions}
L.S., S.C., M.R., K.A. conceptualized the work; E.E. conducted the experiments; L.S, F.D. and A.M. performed the data analysis; L.S. wrote the original draft; K.A. and M.R. reviewed and edited the manuscript. 
\nolinenumbers

\end{document}


\title{\large Supplemental Materials: Decision-making in light-trapped slime molds involves active mechanical processes}

\author{Lisa Schick}
\affiliation{%
 Technical University of Munich, TUM School of Natural Sciences, Department of Bioscience, Center for Protein Assemblies (CPA), Germany
}
\author{Emily Eichenlaub}
\affiliation{%
 Technical University of Munich, TUM School of Natural Sciences, Department of Bioscience, Center for Protein Assemblies (CPA), Germany
}
\author{Fabian Drexel}
\affiliation{%
    Technical University of Munich, TUM School of Natural Sciences, Department of Bioscience, Center for Protein Assemblies (CPA), Germany
}
\author{Alexander Mayer}
\affiliation{%
    Department of Mathematics, UCLA, Los Angeles, California, USA
}
\author{Siyu Chen}
\affiliation{%
    Max Planck Institute for Dynamics and Self-Organization, Göttingen, Germany
}
\author{Marcus Roper}
\affiliation{%
    Department of Mathematics, UCLA, Los Angeles, California, USA
}
\author{Karen Alim}%
 \email{k.alim@tum.de}
\affiliation{%
 Technical University of Munich,  TUM School of Natural Sciences, Department of Bioscience, Center for Protein Assemblies (CPA), Germany
}%

\date{\today}

\maketitle
\widetext
\section{Calculation of characteristic quantities}
\subsection{Relative axis length}
\begin{figure}[h!]
    \centering
    \includegraphics[width=8.6cm]{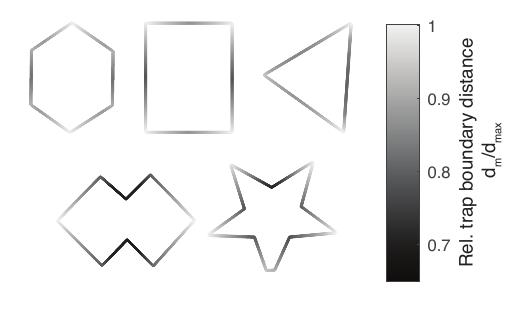}
    \caption{Relative axis length for all explored shapes.}
    \label{suppFig:bdist}
\end{figure}

\noindent We introduce the relative axis length $\frac{d_\mathrm{m}}{d_\mathrm{max}}$ to parameterize a fans exit location along the trap boundary. Here, $d_\mathrm{m}$ represents the longest possible internal path through the shape at the exit point, normalized by $d_\mathrm{max}$, the longest axis of the entire shape. This ratio quantifies how closely the fan’s escape aligns with the longest axis of the shape (Fig.~\ref{suppFig:bdist}).

\newpage
\subsection{Phase analysis for all shapes}
\begin{figure*}[h!]
    \centering
    \includegraphics[width=17.4cm]{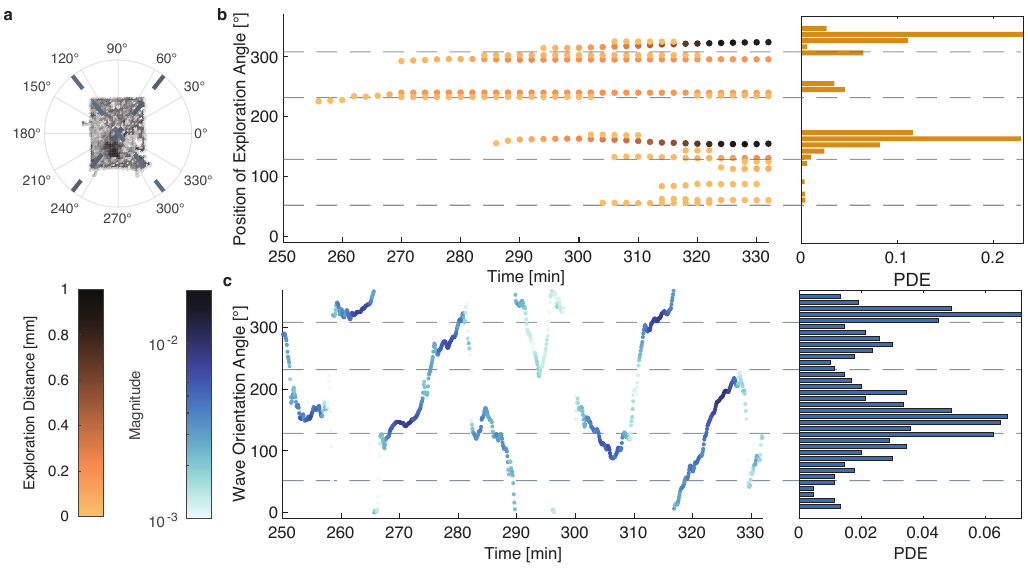}
    \caption{During exploration, the angular fan location aligns with the orientation of the peristaltic wave. (a) Orientation of longest axes (gray dashed lines) within the square-shaped trap. (b) During the exploration phase, the angular component of the center of mass of localized fans over time is defined as the trap being completely filled and lasting until escape. Fans cluster around the longest axes of the trap but emerge elsewhere as well. Exploration distance is measured using a minimal Euclidean distance of the fan center to trap the boundary. (c) Orientation of peristaltic wave over time mainly follows longest axes of shape with gradient magnitude being especially large and wave more stable along the escape axis of $~130^\circ$. Histograms in (b) and (c) display the relative probability (PDE) of the event count. The gray dashed lines indicate the angles of the longest axes within the square-shaped trap.}
    \label{suppFig:square}
\end{figure*}

\begin{figure*}[h!]
    \centering
    \includegraphics[width=17.4cm]{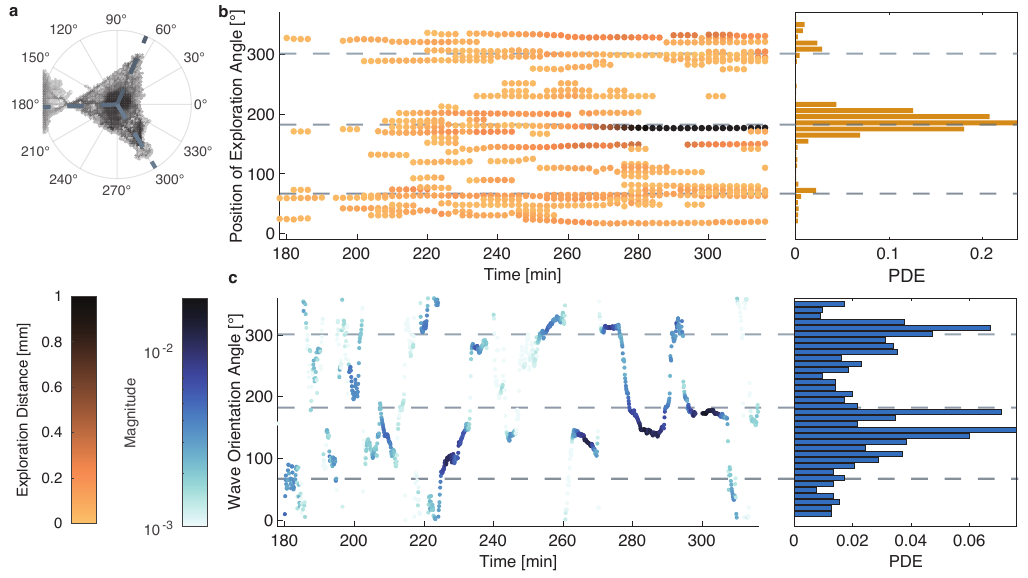}
    \caption{During exploration, the angular fan location aligns with the orientation of the peristaltic wave. (a) Orientation of longest axes (gray dashed lines) within the triangular-shaped trap. (b) During the exploration phase, the angular component of the center of mass of localized fans over time is defined as the trap being completely filled and lasting until escape. Fans cluster around the longest axes of the trap but emerge elsewhere as well. Exploration distance is measured using a minimal Euclidean distance of the fan center to trap the boundary. (c) Orientation of peristaltic wave over time mainly follows longest axes of shape with gradient magnitude being especially large and wave more stable along the escape axis of $~180^\circ$. Histograms in (b) and (c) display the relative probability (PDE) of the event count. The gray dashed lines indicate the angles of the longest axes within the triangular-shaped trap.}
    \label{suppFig:triangle}
\end{figure*}

\begin{figure*}[h!]
    \centering
    \includegraphics[width=17.4cm]{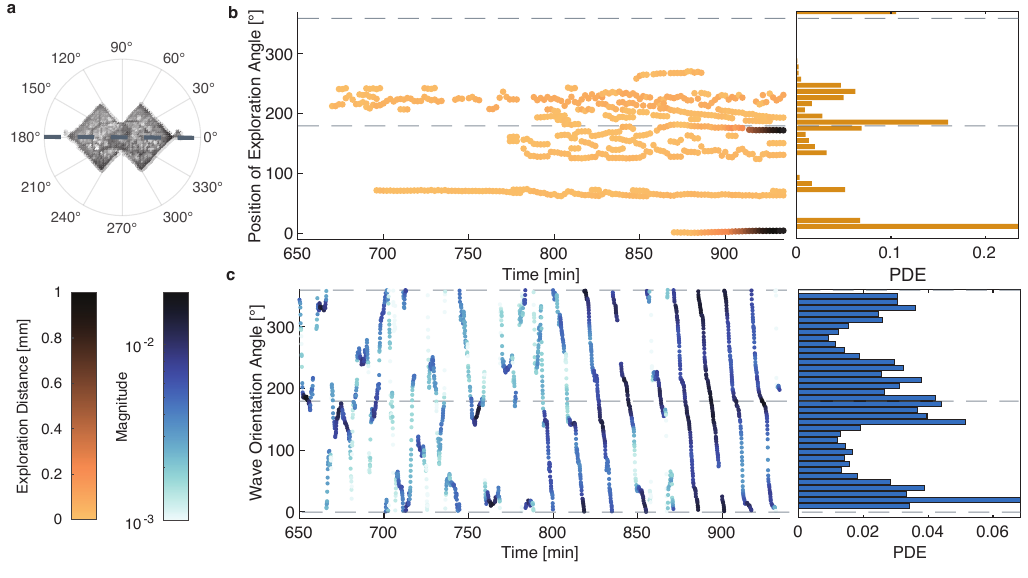}
    \caption{During exploration, the angular fan location aligns with the orientation of the peristaltic wave. (a) Orientation of longest axes (gray dashed lines) within the zigzag dragon-shaped trap. (b) During the exploration phase, the angular component of the center of mass of localized fans over time is defined as the trap being completely filled and lasting until escape. Fans cluster around the longest axes of the trap but emerge elsewhere as well. Exploration distance is measured using a minimal Euclidean distance of the fan center to trap the boundary. (c) Orientation of peristaltic wave over time mainly follows longest axes of shape with gradient magnitude being especially large and wave more stable along the escape axis of $~180^\circ$. Histograms in (b) and (c) display the relative probability (PDE) of the event count. The gray dashed lines indicate the angles of the longest axes within the zigzag dragon-shaped trap.}
    \label{suppFig:polygon}
\end{figure*}

\noindent At the onset of exploring the trap's surroundings, the plasmodium covers the entire blue light-free region inside the trap, extending and retracting fans in multiple directions. Over approximately 1.5~h leading up to escape, fans briefly push into the blue light before retracting. Here, we display the exploration for the square (Fig.~\ref{suppFig:square} b, Supplemental Movie~3), triangle (Fig.~\ref{suppFig:triangle} b, Supplemental Movie~4) and zigzag-diamond (Fig.~\ref{suppFig:polygon} b, Supplemental Movie~5). 
Fan growth remains visible all over the shapes, clustering around the longest axes. Furthermore, the peristaltic wave direction circulates over the shapes aligning with the exploration. More stable orientation directions indicated by high gradient magnitude (dark blue) orient along the longest axes of the individual traps. 
\newpage
\newpage

\section{Near-circular trap shapes}
\begin{figure}[h!]
    \centering
    \includegraphics[width=17.4cm]{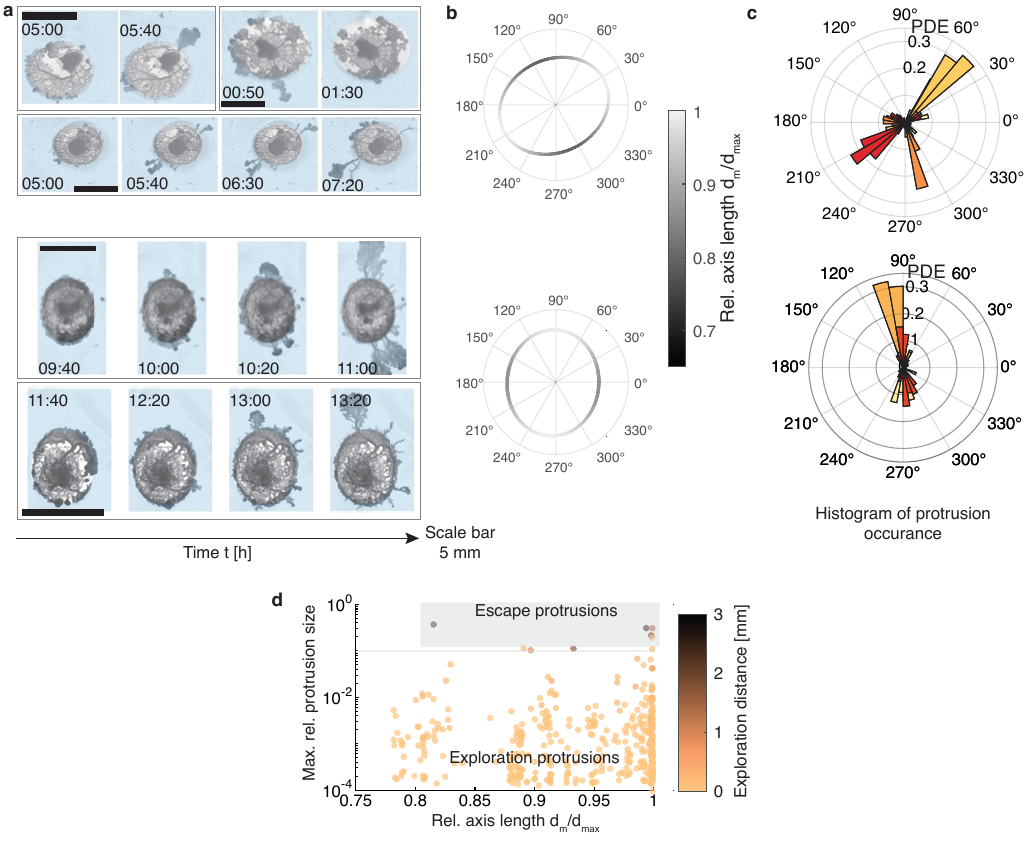}
    \caption{Longest axis alignment guides escape despite near-circular confinement. (a) Five individual specimen growing in near-circular traps with their longest axis oriented [30$^\circ$, 210$^\circ$] (b, top) or [90$^\circ$, 270$^\circ$]) (b, bottom). (c) Histograms of exploration frequencies for the different datasets in varying yellow-red shades) corresponding to the different orientations indicate escapes oriented similarly to axis orientation of trap. (d) Small protrusions continue to emerge all around the trap boundary (exploration protrusions), yet escapes only happen close to the longest axis within the shape (escape protrusions highlighted by gray box and 100\% opacity). Exploration distance: Euclidean distance of protrusion center of mass to protrusion exit point on trap boundary.}
    \label{suppFig:circleData}
\end{figure}

    \noindent In polygonal confinement, escape locations exhibit a strong correlation with the longest axis of the shape, which corresponds to regions of maximal curvature of the trap boundary. However, we also observe escapes in nominally circular confinements (see Fig.~\ref{suppFig:circleData}a). Due to optical aberrations, the circular traps deviate from perfect circularity and are therefore slightly ellipsoidal. Yet, variations in curvature and axis length are significantly lower than in the other polygonal geometries (see Fig.~\ref{suppFig:circleData}b). This prompts the following question: How does \textit{Physarum polycephalum} select escape locations in the absence of a dominating longest axis? Addressing this question experimentally is challenging because defining directionality in near-circular geometries is difficult. Furthermore, testing for uniform escape distributions would require substantially larger sample sizes. In addition, escape protrusion selection by \textit{P.~polycephalum} can be biased by protrusions reaching the outer boundary of blue light confinement, giving preference to those irrespective of confinement shape. Despite these constraints, analysis of protrusion orientations reveals a clear preference for the direction along the longest axis in near-circular shapes (see Fig.~\ref{suppFig:circleData}c, individual data sets shown in varying yellow–red shades), although outbreaks are not exclusively confined to this direction. It is noteworthy that exploratory protrusions remain in all directions. This phenomenon is further substantiated by the exploration versus escape analysis (Fig.~\ref{suppFig:circleData}d). Collectively, these observations indicate that, even in geometries with weakly defined longest axes, \textit{P.~polycephalum} exhibits biased exploration along directional cues while concurrently maintaining broader exploratory activity that is not strictly axis-dependent. 
\newpage
\section{Transport efficiency in closed networks of contracting tubes}
\noindent For a network of tubes with volume change rates $\dot{\vec{V}}=\frac{\partial \vec{V}}{\partial t}$ enumerated for each network tube in vector notation, the mean volumetric flow in each tube $\vec{Q} = \mathbf{\Gamma} \dot{\vec{V}}$ is determined by a linear mapping $\mathbf{\Gamma}$ that encodes network topology and geometry via Kirchhoff's laws~\citep{wilkinson_Flow_2023}. Total viscous dissipation comprises pressure-driven flow through the network and local extensional flow within tubes:
\begin{equation}
    D_{\mathrm{tot}} = \underbrace{\dot{\vec{V}}^\top \mathbf{\Gamma}^\top 
    \mathbf{K}^{-1} \mathbf{\Gamma} \dot{\vec{V}}}_{D_{\mathrm{transport}}} 
    + \underbrace{\dot{\vec{V}}^\top \mathbf{K}^{-1} \dot{\vec{V}}}_{D_{\mathrm{local}}},
\end{equation}
where $\mathbf{K}$ is the diagonal conductance matrix. We define transport efficiency as:
\begin{equation}
    \Omega = \frac{\dot{\vec{V}}^\top \mathbf{P}^\top \mathbf{\Gamma}^\top 
    \mathbf{K}^{-1} \mathbf{\Gamma} \mathbf{P} \dot{\vec{V}}}{\dot{\vec{V|}}^\top \mathbf{K}^{-1} \dot{\vec{V|}}}
    \label{eq:efficiency}
\end{equation}
Here, the projection matrix $\mathbf{P} = \mathbf{I} - \mathbf{K} \vec{1} \vec{1}^\top / \mathrm{Tr}(\mathbf{K})$ enforces volume conservation: since no fluid enters or leaves the network, the total rate of volume change must vanish ($\vec{1}^\top \dot{\vec{V}} = 0$), and $\mathbf{P}$ restricts the analysis to this subspace. The contraction patterns that maximize $\Omega$ satisfy the generalized eigenvalue problem:
\begin{equation}
    \mathbf{P}^\top \mathbf{\Gamma}^\top \mathbf{K}^{-1} \mathbf{\Gamma} 
    \mathbf{P} \, \vec{x} = \Omega \, \mathbf{K}^{-1} \vec{x}
    \label{eq:eigenvalue_problem}
\end{equation}
with eigenvectors representing optimal pumping modes and eigenvalues their efficiencies\ 

\begin{figure}[htbp]
    \centering
    \includegraphics[width=15 cm]{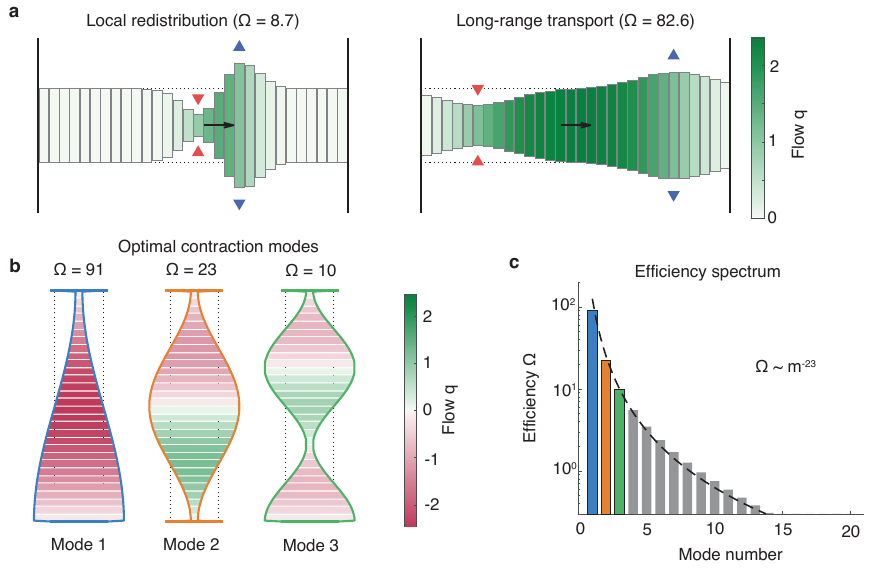}
    \caption{Transport efficiency in a closed tube.
    (a)~Two contraction patterns with equal local extensional flow $D_{\mathrm{local}}$ but disparate transport efficiency. \textit{Left:} Local redistribution ($\Omega = 8.7$). \textit{Right:} Long-range transport ($\Omega = 82.6$). Red/blue triangles: contraction/dilation; color: flow magnitude; arrows: flow direction; dotted lines: resting diameter.
    (b)~First three eigenmodes of Eq.~\eqref{eq:eigenvalue_problem}. Tube boundaries trace contraction amplitude; dotted lines: zero deformation. Fill color: flow direction (green: up; magenta: down). Outline colors match panel~c.
    (c)~Efficiency spectrum decays rapidly with mode number.}
    \label{fig:transport_efficiency}
\end{figure}

\noindent To illustrate, we consider a tube discretized into $n = 30$ independent contractile elements (Fig.~\ref{fig:transport_efficiency}). Figure~\ref{fig:transport_efficiency}a shows two different contraction patterns of same overall local extensional flow $D_{\mathrm{local}}$ but different overall transport efficiency $\Omega$. Local redistribution, with adjacent elements contracting and dilating in anti-phase, produces only short-range flow of low transport efficiency ($\Omega = 8.7$), while long-range transport, contraction and dilation at opposite ends, drives efficient transport flow across the entire tube ($\Omega = 82.6$). The eigenmodes of Eq.~\eqref{eq:eigenvalue_problem} are standing waves of increasing spatial frequency (Fig.~\ref{fig:transport_efficiency}b), with tube boundaries tracing each mode's contraction amplitude. Mode~1 achieves the highest transport efficiency $\Omega$; from mode~1 the spectrum decays rapidly (Fig.~\ref{fig:transport_efficiency}c), note the logarithmic scale, so only the first few modes contribute meaningfully to transport. For higher modes, flow reverses direction along the tube (green vs.\ magenta), creating bidirectional patterns.


